\documentclass[%
 twocolumn,reprint,
 amsmath,amssymb,
 aps
]{revtex4-2}

\usepackage{graphicx}
\usepackage{bm}
\usepackage{bbm}
\usepackage{amsmath}
\usepackage{amssymb}

\begin{document}

\preprint{APS/123-QED}

\title{Spatial Selection and the Multiscale Dynamics of Urban Change}

\author{Jordan T. Kemp$^{1,2,3,4*}$, Laura Fürsich$^{5}$, Elias Tannira$^{3, 4}$, and Lu\'is M. A. Bettencourt$^{3,4}$}
\affiliation{$^1$Institute for New Economic Thinking at the Martin School, University of Oxford, Oxford, OX1 3UQ, UK}
\affiliation{$^2$School of Geography and the Environment, University of Oxford, Oxford OX1 3QY, UK}
\affiliation{$^3$University of Chicago, Department of Ecology and Evolution, Chicago IL, 60637, USA}
\affiliation{$^4$Santa Fe Institute, Santa Fe NM, 87501, USA}
\affiliation{$^5$Linköping University, Institute for Analytical Sociology, Norrköping, 601 74, Sweden}
\affiliation{$^*$corresponding author: Jordan Kemp (jordan.kemp@ouce.ox.ac.uk)}

\begin{abstract}
Demographic and economic growth drive many processes of change in human societies. Yet, standard macroeconomic indicators tend to average over local variations, blurring heterogeneities associated with people's actual choices and agency. To address this gap, we introduce an analytical framework based on levels of selection to decompose economic and demographic growth exactly into endogenous and sorting effects across nested spatial scales. We illustrate this approach with population and income growth in the Chicago metropolitan area. We show that both growth rates and spatial selection effects are most intense at local levels, fat-tailed and spatially correlated. We also find that local selection, a measure of population spatial sorting by income, is concentrated in few spatial units and reflects a clear rebalancing of higher-income populations toward downtown and the semi-periphery, accompanied by the displacement of lower-income residents. At larger scales, local selection relevant for individual decisions partially cancel out, emphasizing the need for more localized analysis of change. 
This approach is general and applies to any other nested complex system such as networks of production, occupations, or innovation enabling the systematic mechanistic understanding of how compositional change relates to growth across scales of organization.
\end{abstract}

\maketitle

\section*{Introduction}

Demographic and economic growth is a fundamental driver of change in modern human societies. While the social value of economic growth is contentious~\cite{jackson2009prosperity}, it is commonly used to measure a society's vitality, its ability to produce and consume goods and services and, over time, to improve its stability and social welfare~\cite{barro1991economic}. 
Beyond economics, variations in growth reveal compositional changes in populations \cite{hairston1970natural,lee2002demographic}, and encode latent features of economic agents' environments~\cite{kemp2023learning}, clarifying the foundations of socioeconomic and demographic dynamics.

Growth and change occur at multiple scales from local communities to cities and nations, amplifiying or mitigating patterns of development and inequality~\cite{cholli2024understanding}. For example, there is typically local population sorting by income leading to stronger economic inequalities between neighborhoods than within them~\cite{bettencourt2025decoding}. Consequently, policy guided primarily by aggregate indicators such as population growth, total wages or national GDP can produce unforeseen distributional effects, especially by (dis)advantaged households. 
Population aggregation in policy studies often averages away local socioeconomic dynamics relevant to individual households~\cite{glaeser1995economic}, undermining any theory of real agents responsive to heterogeneous local change~\cite{kemp2023learning} and creating puzzles in analysis and interpretation. We will show that such local and relative variations in income and population dominate socioeconomic dynamics in developed societies, where overall growth is slow. 

Yet resolving these hidden dynamics requires more than simply increasing spatial resolution. 
It requires understanding how processes observed at different scales compose into aggregate change.
Such multiscalar analyses remain difficult due to several interrelated factors. Data are embedded in sociopolitical contexts and reflect local regulatory environments that vary by local constituency (e.g. municipality)~\cite{oosterhaven2024disaggregating}.
Moreover, empirical growth patterns depend on dynamic processes, such as the spatial sorting of heterogeneous populations, capital, and income dynamics~\cite{maestas2016impact,bloom2003demographic}, making results highly sensitive to different aggregation methods~\cite{hulten2010growth,melitz2015decomposition}.

Multi-scale spatial accounting is therefore essential to studying socioeconomic dynamics in cities~\cite{bettencourt2025decoding,duranton2004micro}.
Several methodological approaches have been developed for this purpose, including hierarchical linear, spatial econometric, and structural decomposition models~\cite{anselin1988spatial,elhorst2017spatial}.
These methods typically sum growth contributions across sectors or spatial units while accounting for spatial and temporal correlations~\cite{rey2021pysal}. Although influential in research and policy, such methods rely on restrictive assumptions such as linearity, separability, or phenomenological fits that limit interpretability~\cite{manski1993identification,helbing2015societal}.
Furthermore, while these tools clarify dynamics, they do not provide a principled way to partition growth into within-unit dynamics and between-unit selection across spatial scales.

\begin{figure*}
 \centering
 \includegraphics[width=\linewidth]{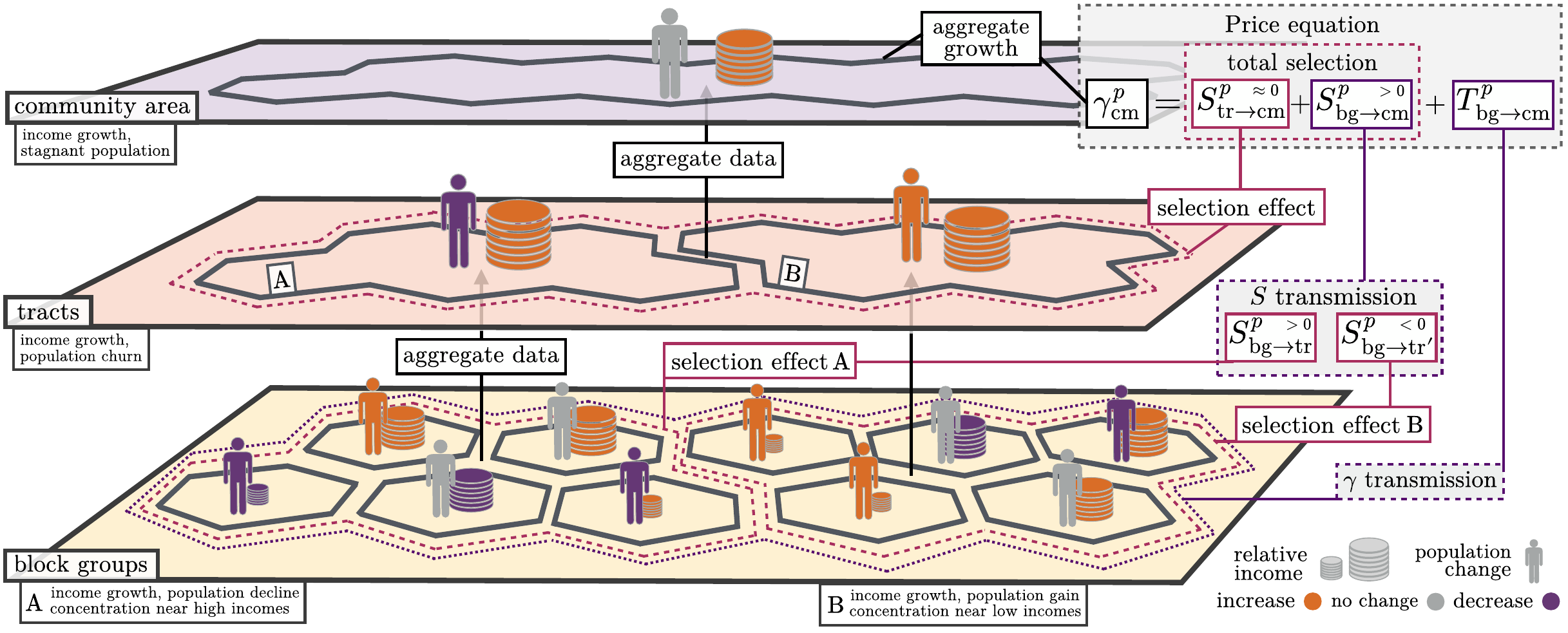}
 \caption{{\bf Schematic of 3-level decomposition in a typical community area with no population growth but positive income growth.} 
 Between tracts (child-level) the population rebalances from the low-income A to the high-income tract B. Between block groups (lowest level), populations and incomes are growing in tract B's lower income areas, and populations are declining in tract A's low income areas. These changes are recorded as selection effects and transmitted via averaging to the community level.}\label{fig:fig2}
\end{figure*}

Resolving heterogeneous dynamics across scales is not a challenge unique to socioeconomic studies. 
Statistical methods in data sciences, ecosystems and economic supply networks often aim to describe multiscalar dynamics of interaction~\cite{chave2013problem} and shock propagation~\cite{carvalho2014micro}.
A unifying feature of these systems is the need for an exact, model-free identity to separate endogenous change within units from differential growth between them.
To this end, the Price equation~\cite{price1972fisher} offers an exact decomposition separating endogenous growth effects from population sorting, or selection.
It is also recursive, making it well suited for measuring variations in subpopulations at successively lower levels. This hierarchical structure \cite{bettencourt2021introduction} captures multi-level selection, enabling the analysis of heterogeneous growth patterns across multiple scales. Here, we measure selection as an emergent property of residential sorting, measuring signatures of preferences under structural constraints such as housing costs, discrimination, zoning; analogous but distinct from segregation measures such as Theil's index and the multigroup index 
\cite{theil1971principles, reardon2002measures}(discussed further APP \ref{selectioninterpretation}). 

Fig. \ref{fig:fig2} illustrates the selection framework in a community with relatively slow aggregate population growth but substantial internal population turnover. The schematic community consists of tracts A and B, which initially have similar populations, income levels, and positive income growth. Within tract A, individuals increasingly sort into high-income block groups, while in tract B, population shifts toward lower-income block groups. If the positive selection in A outweighs the negative selection in B, block groups contribute positive selection on average.
At the tract level, however, A and B maintain similar aggregate income levels. As a result, although the population of tract A declines while tract B grows, selection between tracts remains negligible. 

The Price framework therefore separates and sums selection contributions across spatial units, allowing substantial selection-driven income growth to emerge despite minimal aggregate population change.
Its recursive structure preserves multilevel spatial stratification, addressing a longstanding challenge in urban sociology of linking micro-ecological processes to macro-structural outcomes~\cite{mckenzie_ecological_1925,bettencourt2021introduction}.

\begin{figure*}
 \centering
 \includegraphics[width=\linewidth]{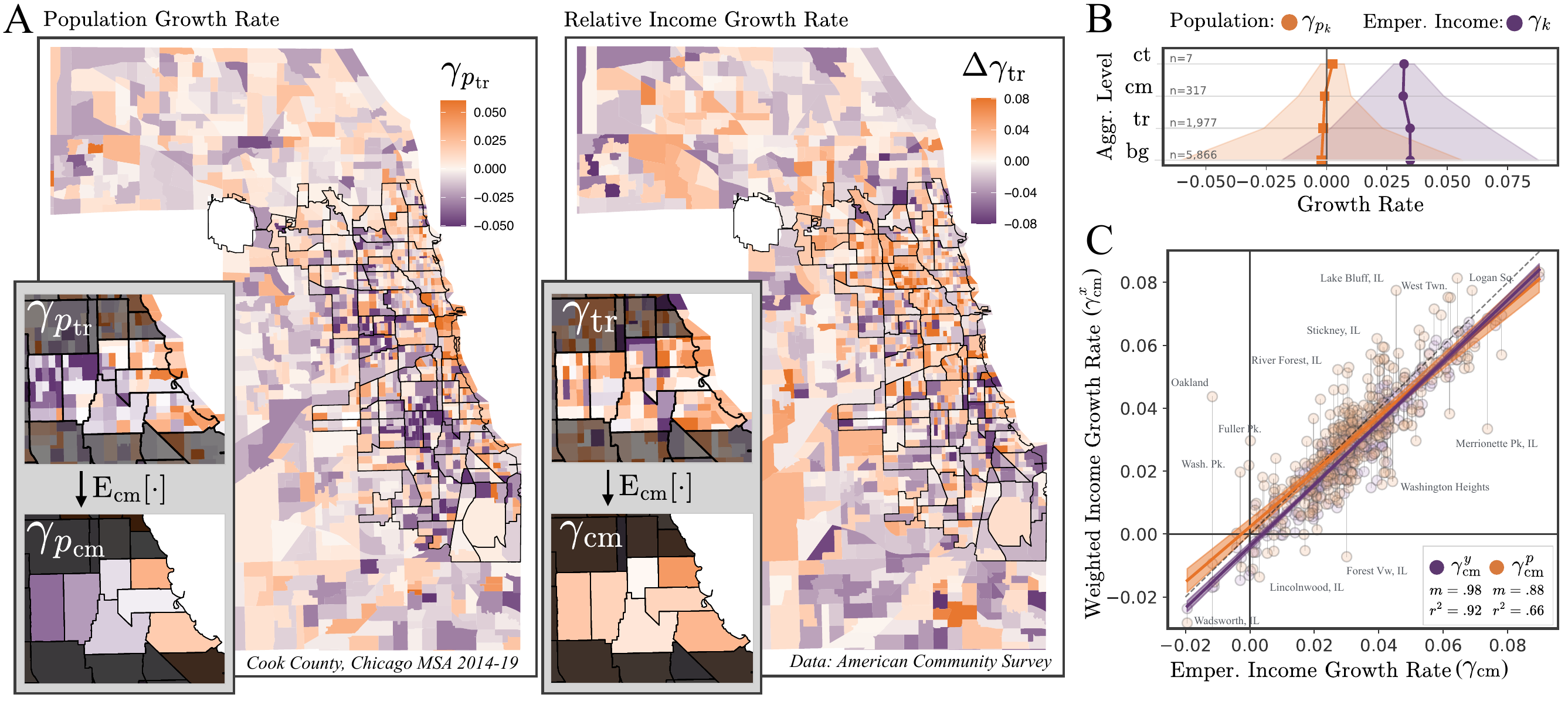}
 \caption{{\bf Local income and population growth rates are heterogeneous and spatially correlated.} In Chicago, A) population growth (left) and relative income growth (right) in neighborhoods (census tracts) vary strongly, on the order of $ \gamma_{p_\textrm{tr}}= \pm 5\%$/year and $\Delta\gamma_{p_\textrm{tr}}=\pm 8\%$/year respectively. The insets show how the local heterogeneity and correlations in growth rates at the tract level wash out under averaging ($\textrm{E}_\textrm{cm}[\cdot]$) into community areas. B) Heterogeneity, measured by the standard deviation at each organizational level across the MSA, decreases under aggregation. Shifts in the mean value encode selection effects to be explored in this paper. C) Differences between income growth aggregation methods, defined in Eqs. \ref{eq:priceaggregation} and \ref{eq:oleypakesaggregation}, and a standard measure of aggregate income growth via Eq. \ref{eq:empiricalgrowth} encode information about growth dynamics.}\label{fig:Fig1}
\end{figure*}

This provides a simple, exact, and general method
to disaggregate change in any socioeconomic quantity in terms of local demographic effects. As an illustration, we explore the rich spatial heterogeneity of urban population and income growth in Chicago, which we show is averaged away in standard aggregate metrics. We then show how the Price equation
decomposes aggregate income growth into selection and
endogenous effects at each organizational scale.
Because the decomposition is exact, scale-explicit, and model-free, it yields interpretable statistics that attribute macroscopic change to processes operating at specific spatial scales.
This enables comparative analysis of growth trends across local
and citywide scales, for instance in response to policy. 
We demonstrate its utility by identifying local
population change effects, such as gentrification and income concentration. 
Then, we use the Price equation,
to aggregate these effects and produce exact summary statistics attributing the macroscopic growth effects to
variations at each level
We end the paper by demonstrating how these neighborhood processes predict future transformations in Chicago.

\section*{Results} 
Chicago's dynamic demographic and economic history have been the subject of over a century of urban research, particularly towards understanding its heterogeneity and neighborhood effects \cite{sampson_great_2012,park_city_1984}. 
Since 1990, the city’s population experienced bursts of local growth and contraction, resulting in a net loss of approximately 40k residents. We report these dynamics for Cook County (the urban core of Chicago) between $t=2014$ and $t^\prime=2019$ through the annualized population growth rate measured across census tracts \textrm{tr}, $\gamma_{p_{\textrm{tr}}}=\ln(p_{\textrm{tr},t^\prime}/p_{\textrm{tr},t})/\Delta t$ (where $\Delta t=5$yr, see App A).
Fig. \ref{fig:Fig1}A shows population growth over this period in the east along the lake shore. This is balanced by inland population decline mostly in the southern and western regions of the city. By contrast, we observe almost stagnant population growth at the MSA level (with annual $.025\%$ growth), visualized in Fig \ref{fig:Fig1}B. 
As demographic growth slows down across the world, local turnover is becoming the dominant source of change common to many other cities and regions.   

Income growth in Chicago over the same period has had a different character. MSA-wide nominal income has grown at a rate of $\gamma_{\textrm{m}}=3.42\%$/year, in line with national patterns (annual average inflation rate over this period was about 1.5\%). The empirical income growth rate is defined as 
\begin{equation}
\label{eq:empiricalgrowth}
\gamma_{j}\equiv\Delta\ln\textrm{E}_j[y_k] =\ln y_{jt^\prime}-\ln y_{jt},
\end{equation}
with comparable amounts of spatial and distributional heterogeneity reported in Fig.~\ref{fig:Fig1}A and B. We report local growth relative to the county in sub-figure A, where $\Delta\gamma_\textrm{tr}=\gamma_{\textrm{tr}}-\gamma_\textrm{ct}$ to identify where income disparities are increasing. Observed diverse patterns of change in terms of income and population motivate examining more granular, neighborhood level data, which more directly reflect individual experiences~\cite{betancur2002politics,hwang2014divergent}. 

To this end, we apply the Price equation. 
We start with the smallest units of analysis (block groups, $p_\textrm{bg}\sim 1,000$) and characterize each unit's contribution to aggregate rates of income change across a hierarchy of nested scales defined by the US Census, including census tracts (tr), communities (cm for Chicago communities and census places), counties (ct), and the city (m). 
Along the way, we measure selection effects and identify typologies of local change.

The Price equation describes the change in an average population trait, here logarithmic incomes $z=\ln y$. Then, income growth rates, $\gamma = \frac{d z}{dt}$, are averages of variations in logarithmic income at lower levels. 
Making these scales explicit, the growth rate is defined as (App A4)
\begin{equation}\label{eq:priceaggregation}
 \gamma_j^p\equiv \textrm E_j[\Delta\ln y_k]=\sum_{k\in j}(f_{kt^\prime} z_{kt^\prime}-f_{kt}z_{kt}),
\end{equation} 
where $f_{kt}=p_{kt}/\sum_kp_{kt}$ is the population share of unit $k$ at time $t$. 
As we will show in the following section, this method decomposes growth in terms of \textit{population selection} effects (hence superscript $p$) in an analysis consistent with the original formulation of the Price equation \cite{price1972fisher,frank2012natural}.
An alternate composition method that composes aggregate growth rates in terms of income shares, $\gamma_j^y$, ~\cite{olley1992dynamics} is provide in supplement.
Fig \ref{fig:Fig1}C demonstrates that while these quantities generally predict the empirical growth rate well, there are some noticeably strong outliers, suggesting different observable selection effects between aggregation methods.

\begin{figure*}
 \centering
 \includegraphics[width=.63\linewidth,angle=270]{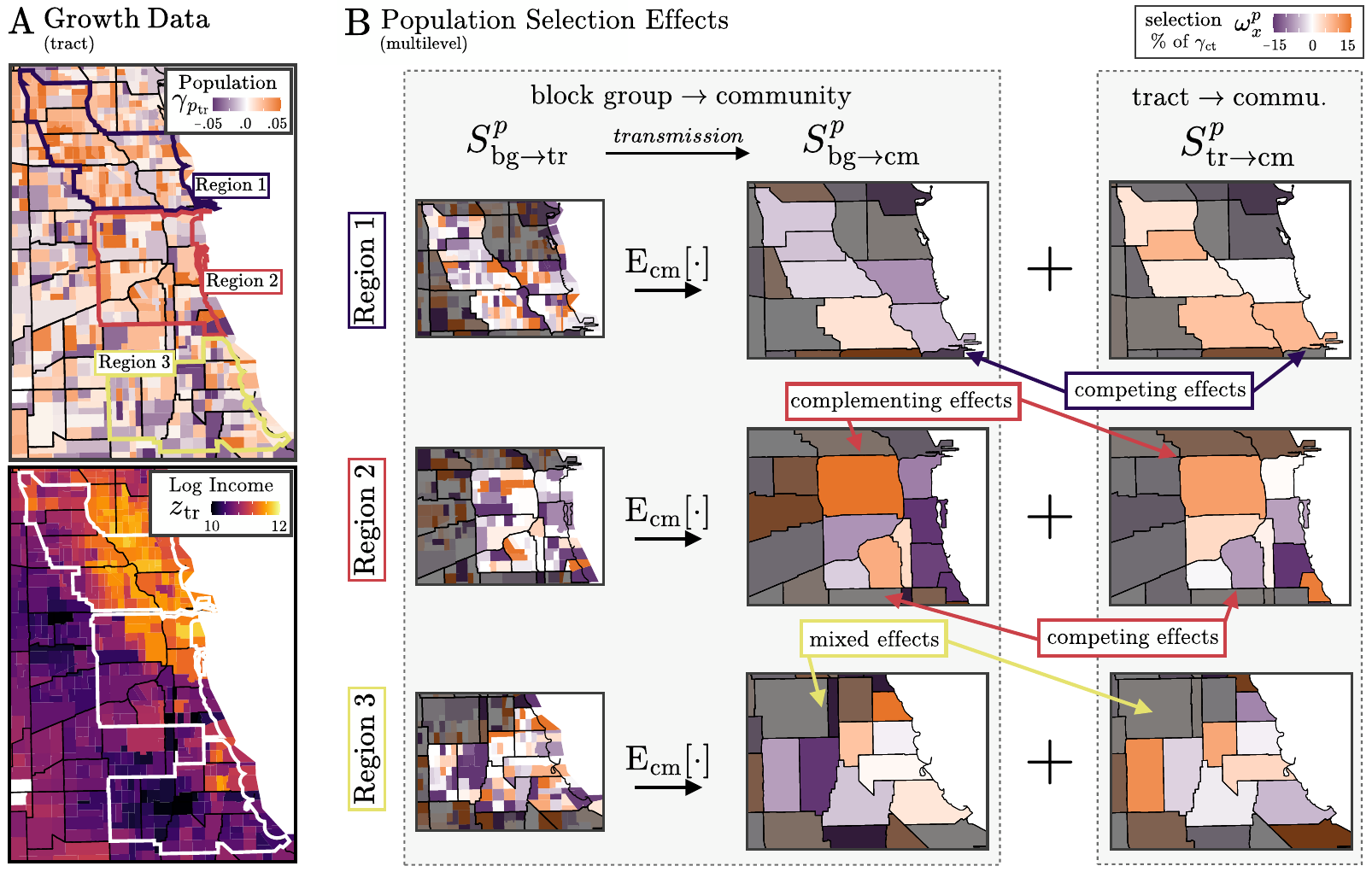}
 \caption{{ \bf Population selection on income varies widely across neighborhoods and scales.} A. The raw income and population growth data are divided into three regions (contour lines). B. Local income selection effects transmitted to the community area level, mapped for the city of Chicago by region are computed from block group-level effects (left) and tract-level effects (right). The aggregate selection effect for each unit is amplified or hidden, depending on whether the selection is complementary (same-sign) or competing (opposite-sign). }
 \label{fig:fig3}
\end{figure*}
\subsubsection*{Multiscale Growth Decomposition}


Though more widely used in biology, the Price equation is a general statistical result accounting for selection at different scales, known as levels of selection~\cite{price_selection_1970,frank2012natural}. 


The aggregation in Eq. \ref{eq:priceaggregation} encodes the local population reweighting over time. This is measured relative to sibling units by $ w_k\equiv \ \exp[\gamma_{p_k}-\gamma_{p_j}]=f_{k,t^\prime}/f_{k,t}$.
, for common parent unit $j$.
In these terms, the Price equation restates Eq. \ref{eq:priceaggregation} as
 \begin{equation}\label{pricemaintext}
 \gamma_j^p=\textrm E_j
 \big[ w_k\gamma_k^p
 \big]+\textrm{cov}_j\big[ w_k,z_k\big].
\end{equation} 
The average growth rate of the higher level unit is equal to the average over the reweighted growth rates, plus the covariance between growth rates and log incomes across all lower level units. The $transmission$, given by the first term $T_{k\rightarrow j}^p\equiv E_j[ w_k\gamma_k^p]$, measures the endogenous contribution of each subunit to the average growth of the larger unit. For example, changes in resident population's income due to raises, job losses, or retirements would belong to this term. $Selection$, given by the second term $S_{k\rightarrow j}^p\equiv\textrm{cov}_j\big[ w_k,z_k\big]$, measures the statistical effects of population sorting (residential moves) between spatial units. 
When selection is positive, population growth rates co-vary with levels of income, so that population grow in richer areas or decline in poorer neighborhoods. Negative selection means that population growth and incomes are anti-correlated, meaning that richer neighborhoods loose population or poorer ones gain it.
This provides us with a two by two typology of local growth, which we explore later. 

The Price equation is naturally recursive~\cite{frank2012natural}, so that the growth rate of unit $j$ is described in terms of the contributions from lower units $k$, and so on, in terms of their subunits $l$.
The precise decomposition method is provided in Appendix \ref{section:pricedecomp}.
The result states that the growth rate is equal to the sum over the selection effects, measured independently and locally at each level, transmitted to the aggregate level through population-weighted averaging of subunits. 

The average growth rate for the MSA can then be disaggregated in terms of the smallest subunits, $ \textrm{bg}$, and the straightforward summation of three intermediate levels, tr, cm, and ct as
 \begin{equation}
\begin{split}\label{priceequationmultilevelmaintext}
 \gamma_{\textrm{m}}^p=&S_{\textrm{ct}\rightarrow \textrm{m}}^p+S_{\textrm{cm}\rightarrow \textrm{m}}^p
 +
 S_{\textrm{tr}\rightarrow \textrm{m}}^p +S_{\textrm{bg}\rightarrow \textrm{m}}^p
 +T_{\textrm{bg}\rightarrow \textrm{m}}^p
 .
\end{split}
\end{equation} 
Here, local fitness is defined relative to the MSA, $w_k=\bar w_k/\bar w_{\textrm m}$. 
This expression measures the cumulative selection, measured between units grouped within the organizational hierarchies, aggregated to the level of the MSA. 

Figure~\ref{fig:fig3} presents the Price equation defined for several communities using real income and population growth data from Chicago’s urban core. 
For interpretability, we express selection as a percentage relative to the county-wide growth rate, $\omega_k^p=S_{ k}^p/\gamma_{\textrm{ct}}^p*100$. 
Figure~\ref{fig:fig3}A highlights three spatial regions with consistent selection effects, while Fig.~\ref{fig:fig3}B shows their internal selection at the tract level (left), and within tracts across block groups (right).
Purple represents negative selection and orange positive.
In Region 1 (near north/northwest), selection is negative at the block-group level but positive across tracts, indicating net sorting toward high-income tracts but low-income blocks within. This mismatch between scales is evidence of competing selection pressures, where the aggregate pattern is consistent with residents seeking tract-level amenities while being constrained to lower-cost blocks within.
In Region 2, encompassing much of the near-west communities, selection is aligned across scales: individuals consistently sorted into high-income tracts and block groups inland, whereas in the downtown and near-south, sorting favored lower-income units.
Elsewhere, including most of Region 3, cross-scale selection is inconsistent, reflecting more complex, or random, sorting patterns.

\begin{figure*}
 \centering
 \includegraphics[width=\linewidth]{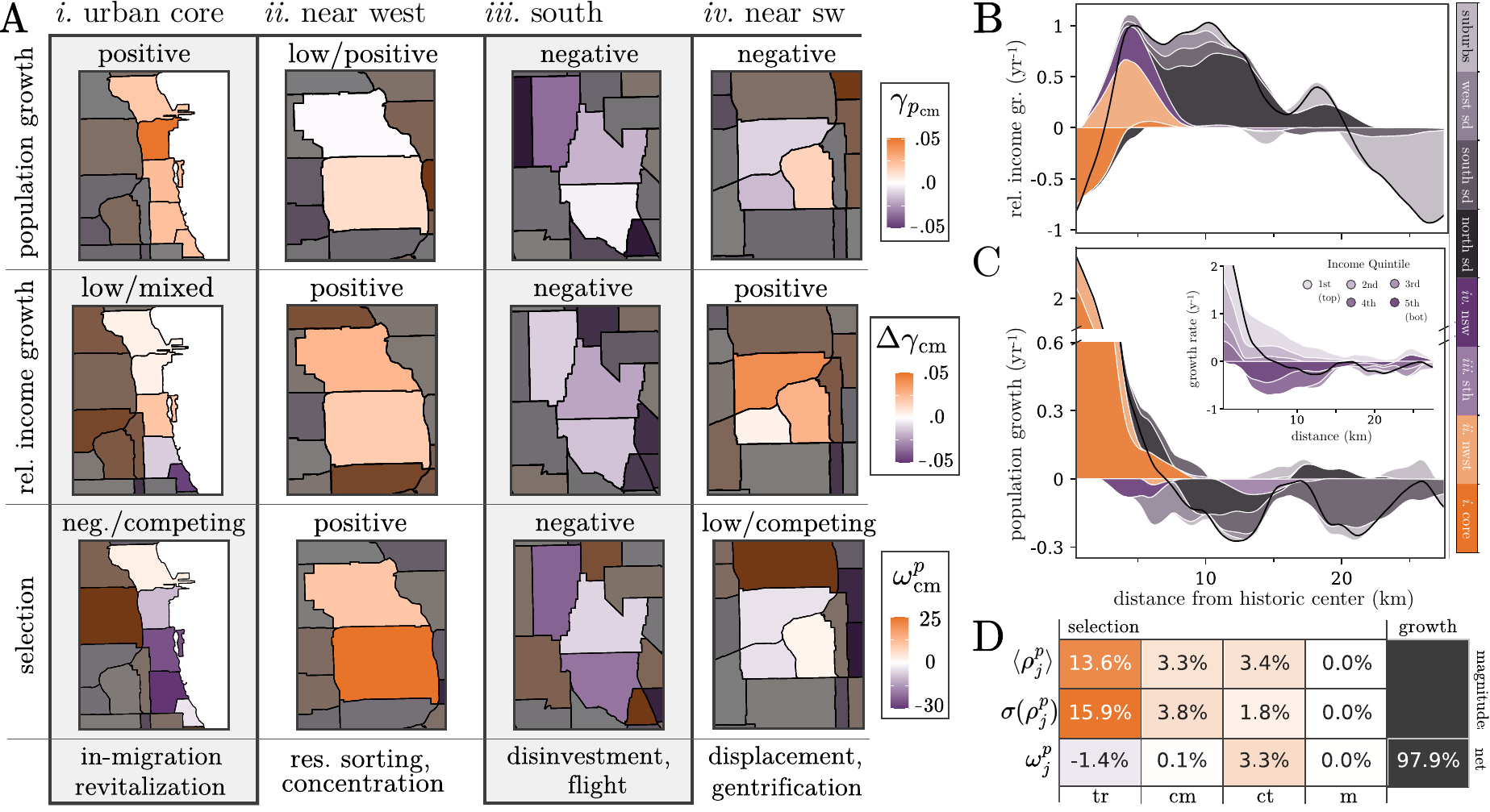}
 \caption{{\bf Analyses of local selection enables a more systematic understanding of urban dynamics}. A. Aggregate income and population growth data are complemented by selection, suggesting potential hypotheses of local change. B. Incomes grew fastest in the semi-periphery, dominated by groups $ii.,$, $iv$, and the north side. C. Population growth was strongest in the urban core, driven by groups $i.$, $ii.$. populations declined in the semi-periphery and periphery. Inset: Middle and high income quintiles drove most of the population growth, replacing low and lower middle income groups. D. Selection magnitudes, $\rho_j^p$, are the highest within tracts, however selection effects transmitted via the Price equation, $\omega_j^p$, are the most consistent across communities in a county. Low transmitted selection effects indicate they largely counterbalance in the Chicago MSA.}
 \label{fig:fig4}
\end{figure*}

\subsubsection*{Data-driven Hypotheses of Urban Change}

Chicago's distinct regions have been the model for studying neighborhood effects, including human development among groups in differing socioeconomic contexts~\cite{park_city_1984,wilson_truly_2006,sampson_great_2012}. For example, urban sociology has emphasized how more powerful groups (by income or race/ethnicity) often promote physical separation from those they deem undesirable~\cite{charles2003dynamics, massey1990american}. Spatial assimilation theory posits that individuals often use socioeconomic resources to access neighborhoods matching their preferences~\cite{massey1994migration}.
Both processes reflect nonrandom population sorting across neighborhoods, which can appear as causal neighborhood effects on various local outcomes from crime to child health \cite{sampson2008moving,sampson2002assessing}.
Our analysis isolates the effects of these selection processes across scales and measures signatures of preferences expressed in structural contexts.

Fig. \ref{fig:fig4}A combines local multilevel selection, $\omega_\textrm{cm}^p=(S_{\textrm{bg}\rightarrow \textrm{cm}}^p+S_{\textrm{tr}\rightarrow \textrm{cm}}^p)/\gamma_\textrm{ct}^p$ to identify where residential decisions tangibly alter the landscape of incomes and inequality. This shows how segregation, gentrification, or suburbanization reinforce or offset one another. 

We observe a rich typology of situations. For example,
in case $i.$ (more affluent lakefront communities) population growth was high but with mixed effects on relative income.
 The mixed-to-negative income selection indicates that much of this migration is into low-income subareas, suggesting potential revitalization of a once exclusive rich area of the city. Case $ii$. shows low population gains with relatively high income growth, coupled with strongly positive selection. This is consistent with income concentration, where higher-income populations disproportionally sort into higher-income areas.
These dynamics contrast with $iii.$ and $iv.$ where populations are stagnating or declining. 
In $iii.$ (lower income southern communities), we observe relative income loss coupled with negative selection, signaling out-migration of relatively higher income residents. This is a pattern commonly discussed with the creation of concentrated poverty~\cite{sampson2012great,bettencourt2021introduction}. Lastly, we observe income gains in some southwest neighborhoods in $iv$, along with competing selection but population stagnation. This suggests a more spatially complex, scale-dependent replacement of incumbent residents. Here, we observe micro-segregation inside affluent tracts: households move into locally high-income tracts (positive between-tract selection) but concentrate in their lower-income block groups (negative within-tract selection).

Figures \ref{fig:fig4}B and C show how local selection contributes to Chicago's evolving demographic landscape. 
Using Eq. \ref{eg:oleypakessupp}, we compose the aggregate growth radially from the zero-point of Chicago's street grid (black curve), in terms of the positive and negative contributions of the four community groups, the three other canonical zones of Chicago (north, south, west), and the suburbs. 

We find that income growth is highest between downtown and the periphery, and is dominated by groups $ii$., $iv$, and the north side. Comparatively, incomes in the core, suburbs, and south side are lagging. Subfigure C demonstrates population growth shares are dominated in the urban core by $i$, $ii$, and the north and south sides. 
Most other regions, including $iii.$, $iv.$, and large swaths of the south and north sides are losing population. 
The inset shows population gains throughout the city are due to middle and high income residents, whereas the lowest income residents are fleeing.
Combined, these diagrams paint a clear typology of change for Chicago marked by economically integrated densification, the replacement of low income residents with a more affluent semi-periphery, and a depopulating periphery.

In the supplement, we measure the amount of overall selection variation, reported as $\rho_{\textrm{tr}}^p=|S_{\textrm{bg}\rightarrow\textrm{tr}}^p|/\gamma_\textrm{Chi}^p$. We find that selection is fat tailed, spatially clustered, as only 20\% of tracts (communities, $\rho_\textrm{cm}^p$) account for 59\% (64\%) of all selection. 
While tract and community-level selection are concentrated similarly, Figure S2A shows that tract-level selection has wider variation and skew than community-level selection, indicating stronger local heterogeneity.
Furthermore, scaling analyses in Appendix \ref{scaling} suggest that population centers have stronger selection dynamics.
Fig. \ref{fig:fig4}D reports both the population-weighted average, $\langle\rho_j^p\rangle$ and its standard deviation, $\sigma(\rho_j^p)$ across spatial units, revealing the largest heterogeneity at the tract level at $13.6\%$, with standard deviation of $15.9\%$.

Despite this local richness, most variation vanishes when aggregated to the MSA level.
The lower panel of Fig. \ref{fig:fig4}D reports the empirical selection effects, $\omega_j^p$, where opposing signs among sibling units offset. 
Selection effects within tracts changes aggregate growth by only $-1.4\%$ while selection across communities contributes $3.3\%$. Overall, selection effects within communities and across the MSA nearly balance, contributing an additional 2.1\% increase to aggregate MSA income growth. 
These results are consistent with scale-sensitive segregation theory \cite{reardon2008geographic}, which finds population sorting is most acutely local.

By treating spatial units as evolving subpopulations under selection, we measure both how prevailing incomes shape spatial selection and how these choices, in turn, influence aggregate income growth across space and in the aggregate. 
We find that selection effects grow more heterogeneous and pronounced at finer spatial scales.
The densification observed in Figure \ref{fig:fig4} supports this observation, people are sorting regionally into closer, higher income communities and locally into lower-income block groups.
An analogous decomposition applies to Eq. \label{eg:oleypakessupp}, which measures spatial income concentration more directly \cite{olley1992dynamics} and will be explored in future work.

The findings underscore the complexity that models of residential selection must capture to reproduce observed dynamics across scales.
They also provide an exact quantitative framework for disentangling contextual effects from compositional sorting for any trait, not just income, thereby advancing explanatory models of urban change.


\section*{Discussion}
A central challenge for the quantitative social sciences and public policy lies in bridging the gap between the socioeconomic conditions individuals experience and how it appears in aggregate statistics~\cite{reardon2008geographic, sampson2012great,bettencourt2025decoding}. 

We have shown how local aspects of economic and demographic change present different realities, with strong local scale and selection at small scales coexisting with apparent stasis on the aggregate. We have also shown how this decomposition is obtained in general via the statistics of multi-level selection, using the Price equation.

Specifically, we showed that growth selection effects in Chicago are spatially concentrated, fat-tailed, and exhibit scaling. We then demonstrated how a systematic analysis of all neighborhoods in a large and complex city produces local hypotheses for neighborhood dynamics of growth, decline and transformation. We concluded by showing that despite strong local variations and a clear story of local rebalancing, selection effects largely cancel out on larger scales in Chicago, resulting in small statistical effects on aggregate economic growth.
Accordingly, any model of residential selection must increase in complexity with finer spatial resolution; consistent with recent findings on neighborhood selection when information is applied to measure multiscale complexity~\cite{bettencourt2025decoding}. 

Stratified urban data analyzes make explicit the coupling between local structural forces and the compositional evolution of metropolitan areas. Future research could extend this framework by modeling how local institutional structures such as zoning, school district boundaries, and neighborhood-level political organization influence selection across space and scale \cite{crowder2012neighborhood}. 
These structures interact with other demographic characteristics such as race and class to shape residential mobility to produce segregation and concentrated poverty and wealth, or indeed to mitigate such concentrations. 

Evidence shows that, even at similar incomes, Black households face higher risks of downward neighborhood moves, with gentrification and displacement amplifying churn among lower-income renters \cite{crowder2005race, south1997migration, quillian1999migration, desmond2016evicted, massey1994migration}. 
Comparative analyses across metropolitan areas, before and after the COVID-19 pandemic, with varying degrees of fragmentation and segregation, would further clarify how these structural conditions affect the balance between spatial assimilation and place stratification. 

Future research could also explore how local place identities, perceived neighborhood boundaries, and mental maps shape the spatial scale at which residential turnover occurs (e.g., boroughs in NYC, community areas in Chicago, micro-neighborhoods in LA). Prior work \cite{logan1983residents,
krysan2007perceiving} suggests that symbolic and cognitive boundaries, e.g., racial blindspots \cite{krysan_racial_2009}, may interact with formal governance structures to reinforce barriers to upward mobility and neighborhood attainment. This analysis could be extended to examine the bidirectional influence of residential choice and capital allocation through the dynamics of point of interest (POI) dependency networks \cite{yabe2025behaviour}, particularly as climate disasters influence dramatic population churn that persists post-recovery \cite{park2024post}.
Ultimately, such work would inform policy at scales relevant to people and households to reduce inequities in neighborhood access and ensure that the benefits of (urban) growth are more equitably distributed.

Lastly, trait-based, multilevel decompositions can summarize emerging disaggregate trends in hierarchical economic datasets such as production \cite{bacilieri2023firm}, occupation networks \cite{del2021occupational}, and infrastructure \cite{sutradhar2024depopulation}.
Revealing scale-dependent change resulting from shocks such as future pandemics, climate events, or economic shocks will inform benchmarks for future disequilibrium economic \cite{wiese2024forecasting} and infrastructure \cite{sutradhar2024depopulation} models, improving our ability to predict out-of-sample structural responses in multi-scale economic systems.

\section*{Acknowledgements}
We thank Andrew Stier and François Lafond for their discussions and comments on the manuscript.
This work is supported by the Urban Science Lab at the University of Chicago, and the Institute for New Economic Thinking at the University of Oxford.
\subsection*{Author contribution}
JTK and LMAB conceived the research. JTK designed the experiments, produced the figures, and developed the draft. JTK and LF produced the maps. JTK and ET analyzed results. JTK, LF, and LMAB wrote the manuscript.

\bibliography{derivation}

\appendix
\section*{Supplementary Material}

To illustrate the problem, we use data from the city of Chicago. Chicago has a highly dynamic demographic and economic history, and has been the subject of over a century of urban research, especially in terms of its population heterogeneity, and its relation to neighborhood effects \cite{sampson_great_2012,park_city_1984}. 
Between the 1850s and 1950s, the city's population, $p$, expanded rapidly, with growth largely concentrated in the urban core. In the postwar period, major transportation infrastructure especially highways, were built fundamentally reshaping the city, reducing central density and promoting the rapid growth of Western suburbs. The population of the city of Chicago peaked in the 1950s at 3.6 million though its suburban population has continued to increase to about 9.2 million in the metropolitan area. 

\begin{figure*}
 \centering
 \includegraphics[width=\linewidth]{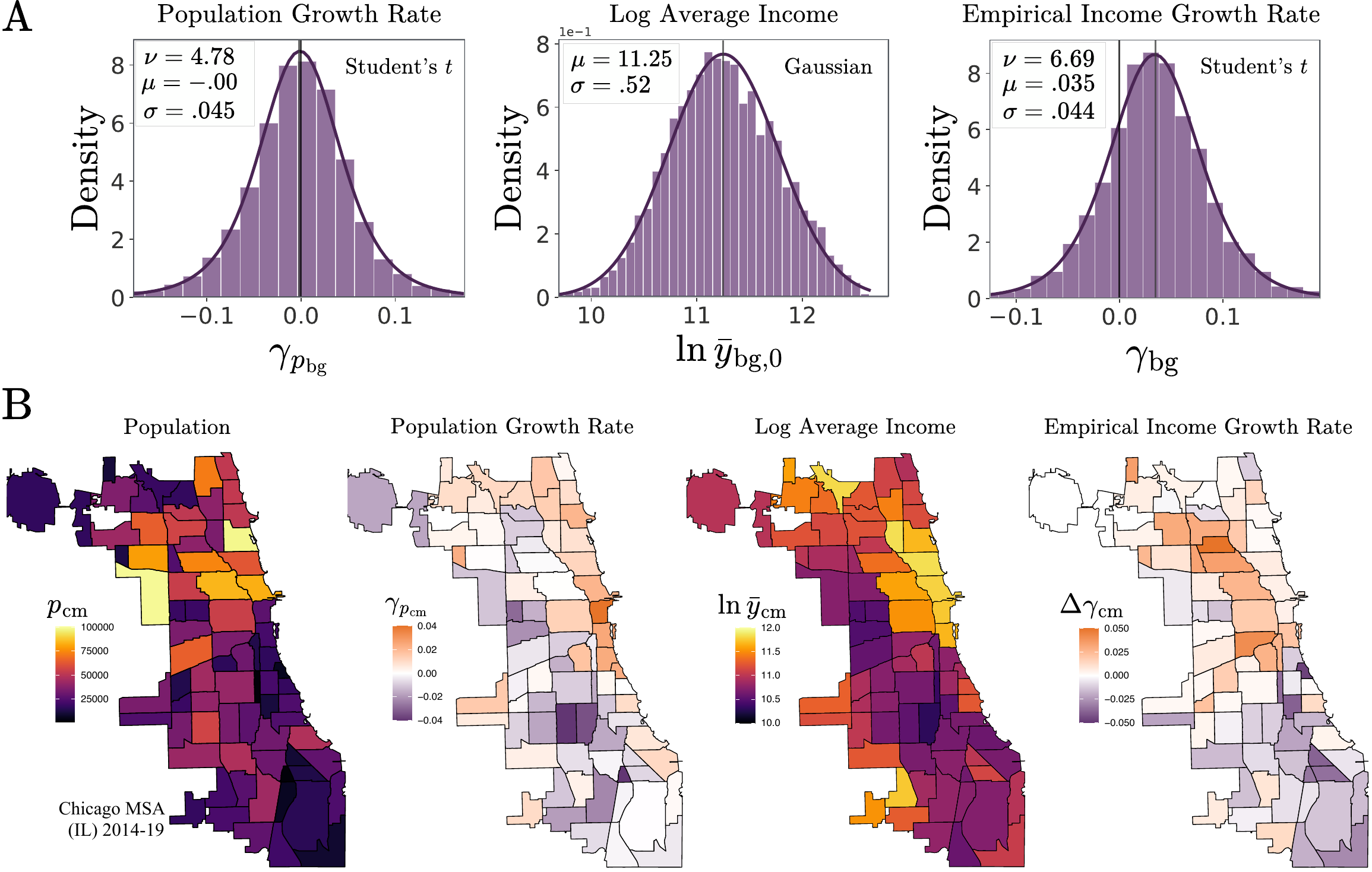}
 \caption{Income and population growth rate data reported between the 2014-2019 period. A) Population and income growth rates are symmetric and fat-tailed, while log incomes are fit with a Gaussian distribution. B) Community-level maps demonstrate that populations were concentrated on the north side and far-west side, however most population growth was measured along the eastern lakefront. Incomes were concentrated along the lakefront in 2014, however most income growth was observed inland in the northwest side neighborhoods, and southwest side communities. }\label{fig:App1}

\end{figure*}

Annual population and income distribution estimates at the census block group ($\textrm{bg}$) level are retrieved from the American Community Survey (ACS). 
We compute changes in population and income as annualized growth rates to build spatial distributions of net migration and income growth. 
We compute population growth rate $\gamma_{p_{\textrm{bg}}}$ for block group $\textrm{bg}$ with population $p_{\textrm{bg}}$, as the log ratio in population between the final and initial year, as $\gamma_{p_{\textrm{bg}}}=(1/\Delta t)\ln(p_{\textrm{bg},t^\prime}/p_{\textrm{bg},t})$, where $t^\prime=2019,t=2014,$ and $\Delta t=t^\prime-t=5$ years. 
We compute these growth rates similarly for larger spatial scales, such as tract (tr), communities (cm, encompassing City of Chicago community areas and census designated places), county (ct), and MSA ($\textrm{m}$), by first summing populations among sibling units in a parent unit (Appendix A2). 
We find that the distribution of population growth rates for all block groups in the MSA is fat-tailed, fitted with a Student-t distribution with mean of $-0.2\%$ (Fig \ref{fig:App1}).
We map the population growth rates at the tract level for a most of Cook County in Fig. \ref{fig:Fig1}A.
 
The definition for the approximate average income of a block group, $ y_{\textrm{bg}}$, is estimated from the midpoint expectation value of ACS income histogram data, where the top income bin is set to $\$400$k (Appendix A).
The annualized growth rate in incomes for a block group is computed similarly as before, $\gamma_{\textrm{bg}}=(1/\Delta t)\ln( y_{\textrm{bg},t^\prime}/ y_{\textrm{bg},t})$.
Nominal income growth rates are also fat-tailed, in agreement with cross-sectional studies of income \cite{guvenen2021data}, and are described by a t distribution with mean $4.78\%$.
For readability, we report growth rates relative to the county as $\Delta\gamma_{\textrm{tr}}=\gamma_{\textrm{tr}}-\gamma_{\textrm{ct}}$, where $\gamma_{\textrm{ct}}=\textrm E_{\textrm{ct}}[\gamma_{\textrm{tr}}]$. 

We show in Fig. \ref{fig:Fig1}B that this rich heterogeneity is lost under aggregation, as the variation in population and growth shrinks as we examine the population statistics at larger levels of aggregation.
We find that the distribution of population growth rates for all block groups in the MSA is fat-tailed, fitted with a Student-t distribution with mean of $-0.2\%$ (Fig \ref{fig:App1}).
We find that the distribution of population growth rates for all block groups in the MSA is fat-tailed, fitted with a Student-t distribution with mean of $-0.2\%$ (Fig \ref{fig:App1}).

\renewcommand{\arraystretch}{1.3}
\begin{table}[h!]
\centering
\begin{tabular}{l p{0.72\linewidth}}
\hline
\textbf{Variable} & \textbf{Name/Description} \\
\hline
$p_j$& Population of spatial unit $j$. \\
$y_{ij}$& Income bin $i$ of unit $j$; when $i$ is excluded, this is the average income over unit $x$.\\
$z_j$& Log average income of unit $j$. \\
$w_j$& Absolute growth of unit $j$. \\
$\bar w_j$& relative growth of unit $j$.\\
$\gamma_j$& Empirical income growth rate of unit $j$.\\
$\gamma_j^p$ & Population-weighted growth rate of unit $j$. \\
$\gamma_j^y$ & Income-weighted growth rate of unit $j$. \\
$\gamma_{p_j}$ & Population growth rate. \\ 
$S_{k\rightarrow j}^p$
 & Selection along incomes aggregated from units $k$ to $j$. 
If the subscript spans more than one level, it is implied that 
$S_{l\rightarrow j}^p = \mathrm{E}_j\!\left[S_{l\rightarrow k}^p\right]$. \\
$\omega_j^p$ & Selection within unit $j$ normalized by the terminal parent’s growth rate. 
In Fig.~3, the terminal parent is Cook County; in Fig.~4, it is the Chicago MSA. \\
$\rho_j^p$& Selection magnitude of unit $j$ normalized by terminal parent growth. 
In Fig.~4, the terminal parent is the Chicago MSA. \\
\hline
\end{tabular}
\caption{Model variables and their definitions.}
\label{tab:variables}
\end{table}
\renewcommand{\arraystretch}{1.0}

\section{Dataset Generation}

To generate the dataset, we read in population and income distribution data from the ACS survey, which is available at the tract level starting in 2010, and available from the block group level between 2014-2023. 
Following the 2020 census, the boundaries of some tracts and block groups were redrawn. 
Tract and block group definitions are therefore not guaranteed to align between 2019 and 2020.
This break conveniently aligns with major productivity and migration shifts during the COVID pandemic in 2020. 
As such, the analysis in this paper is conducted primarily using data between 2014 and 2019. 

\subsection{Data retrieval}

The data provided by the ACS are assigned, at its most granular level, to spatial units called census block groups. Block groups (B), with subscript $\textrm{bg}$ are defined by a FIPS serial code, and its geographic boundaries are by the spatial coordinates of the indices of its shapefile polygons.
We retrieve the population of a block groups using the API code $B01003\_001E$, the number of households with code $B11001\_001E$, and the persons per household as $B25010\_001E$.
The income bin counts in the ACS data are computed by household, so we estimate population using unit-wide household estimates, denoted $h_\textrm{bg}$, weighed by the number of people per household, $\rho_{\textrm{bg}}$
The population is computed as the product of these quantities, $p_{\textrm{bg}}=h_{\textrm{bg}}*\rho_{\textrm{bg}}$.
For block groups with zero population, which could be the case for industrial corridors, bodies of water, or airports, we manually assert $p_{\textrm{bg}}=0$ (in these cases, the persons per household data is set by the Census Bureau to a sentinel value of -666\ldots).
In aggregate we find generally negligible qualitative differences between the distribution of raw population counts, and population counts computed from density adjusted household counts.

We could similarly read in such values at the tract level, denoted in this text as $\textrm{tr}$ (T), the county level, $\textrm{ct}$ (C), or the state level, $\textrm{st}$ (S). 
However, to ensure that population and income counts are consistent across scales, we aggregate such values directly from the block group level in a process described in the following section.
We introduce an intermediate level of aggregation, the community, denoted $\textrm{cm}$, a unit of aggregation reported as community areas for the City of Chicago, and as census designated places in non-Chicago municipalities throughout the state.
The spatial boundaries at this level of aggregation are not standardized in ACS data such that tracts may span more than one community area, and census places span more than one county. 
We match tracts to the community in which lies the centroid of the tract.
Communities that span more than one county at partitioned along the counties' boundaries, so as to partition the population dynamics generally between counties. 
For example, Naperville, IL, which spans DuPage and Will County, is partitioned into "NAPERVILLE (DUPAGE)" and "NAPERVILLE (WILL)" in the data. 
Here, we study data from the Illinois-side of the Chicago MSA, which includes Chicago's 77 community areas, as well municipalities throughout Cook, DuPage, Kane, McHenry, Will, DeKalb, and Kendall counties. 

We read in the income bin data at the block-group level, from ACS codes $B19001\_0XXE$, where $XX$ runs from 02 to 17. 
We assign the income value of the bin to the midpoint of the range the bin spans.
We set the midpoint of the highest income bin, $>\$250K$, to be $\$400K$.
Generally, this value is arbitrary, but is selected to balance between a good fit of income distribution data under a $\$300K$ midpoint, and empirically fat tail of incomes given by $\$500K$.
The effects studied in this paper are determined mostly by the rank-order of incomes, rather than the length of the tail.
Therefore, estimating a precise tail isn't critical to the results, so a precise estimate is beyond the scope of this work. 
For every block group, denoted $\textrm{bg}$, we compute the average income for every spatial unit. 
Consider incomes $y_i $ in bins of population $p_i$ reported in number of people for spatial unit $k$. 
\begin{equation}
 y_{\textrm{bg}}=\sum_i\frac{p_{\textrm{bg},i}y_{\textrm{bg},i}}{\sum_jp_{\textrm{bg},j}},
\end{equation}
where the sum is taken over income bins indexed by $i$. 

\subsection{Aggregation}

For all block groups bg in tract tr, which are matched using the components of the full 11-digit serial code FIPS code (organized by SSCCCTTTTTB), the population is aggregated $\textrm{br}\rightarrow \textrm{tr}$ as
\begin{equation}
 p_{\textrm{tr}}=\sum_{\textrm{bg}\in \textrm{tr}}p_{\textrm{bg}}, 
\end{equation}
where belongingness is determined by common FIPS S, C, and T digits among block groups. 
This aggregation is similarly defined from tracts to communities, $\textrm{tr}\rightarrow\textrm{cm}$, communities to counties as $\textrm{cm}\rightarrow \textrm{ct}$, and from counties to states $\textrm{ct}\rightarrow\textrm{st}$.
Note that community areas are not defined in the FIPS code, and must be identified through the aforementioned geographic matching, both from tracts to community areas, and community areas to counties.
In general, a metropolitan statistical area (MSA) can span more than one state, $\textrm{st}\rightarrow \textrm{m}$.
However, for both simplicity and because the Chicago MSA, the predominant focus of this paper, lies mostly in Illinois, we only aggregate counties within Illinois.

We further define two methods of aggregating incomes, by which we average among children within a spatial unit, and then introduce three methods for aggregating growth rates.
For quantity $x$, we compute the expectation value as the population-weighted average across spatial units within a parent unit of aggregation as $E_j[x_k]=\sum_{k\in j}(p_{k}/p_j)x_k$. 

Here, we study the statistics of both incomes and log incomes. 
We compute the income of a parent unit from its children in two ways. 
First, to aggregate the income in the most standard way, here considered the empirical income, we take the expectation value of incomes across its children (equivalent the expectation value of the sum over child bins). 
This is defined as 
\begin{equation}\label{aggincome}
 y_j\equiv\textrm E_j[y_k]=\sum_{k\in j}f_{k|j} y_k.
\end{equation}
We also aggregate the log income as
\begin{equation}\label{logincome}
 z_j=\textrm E_j[\ln y_{k}]=\sum_{k\in j}f_{k|j}\ln y_k
\end{equation}
This definition differs from the log of Eq. \ref{aggincome}, up to Jensen's inequality, and is not a standard method for aggregating incomes or growth rates. 
However, we show that this aggregation encodes informative population dynamical data, measured through changes in population.

\subsection{Population growth statistics}
We measure the change in population of a block group between two years $t<t^\prime$ with the compound growth rate over the time period.
From the number of households, the population growth rate is computed over the interval $\Delta t=t^\prime-t$ as 
\begin{equation} \gamma_{p_{\textrm{bg}}}\equiv \frac{1}{\Delta t}\ln\frac{ p_{\textrm{bg},t^\prime}}{ p_{\textrm{bg},t}}=\frac{\ln p_{\textrm{bg},t^\prime}-\ln p_{\textrm{bg},t}}{\Delta t}=\frac{\Delta\ln p_{\textrm{bg}}}{\Delta t}.
\end{equation}
\noindent This provides an intuitive measure for population change across arbitrary timescales. 
The \textit{empirical growth rate} is computed for higher levels of aggregation by first aggregating incomes, then computing growth rates. 
The growth rates in incomes across Chicago city proper are visualized in \ref{fig:Fig1}A. 
We find that, although the population growth rate across all of Cook County, converted from households, is around $-.025\%$, we see from the figure that the real changes are highly heterogeneous. 

We can also express growth in terms of the population relative to an earlier time step, and is simply the exponential of the growth rate. 
It is a positive semi-definite quantity defined
\begin{equation}
 w_{\textrm{bg}}=\frac{p_{\textrm{bg},t^\prime}}{p_{\textrm{bg},t}}.
\end{equation}
We compute the relative growth across sibling spatial units by taking the ratio of its growth to its $n$th parent.
For example, we compute the growth of a block group relative to all other block groups in its county by computing
\begin{equation}
 \bar w_{\textrm{bg}}=\frac{w_{\textrm{bg}}}{w_{\textrm{ct}}}=\textrm{exp}[\gamma_{\textrm{bg}}-\gamma_{\textrm{ct}}]
\end{equation}

There is notational ambiguity as to which value a blockgroup is normalized, so it will always be specified in text. 
These two measures of population change will be used throughout the analysis. 

\subsection{Growth statistics}

So far, we have defined income and population aggregation methods, and some statistical measures of population dynamics. 
In this section, we compute growth dynamics for incomes. 
We compute the empirical growth rate for that spatial unit across time span $\Delta t=t^\prime-t$ as the difference in the logarithm of empirical incomes
\begin{equation}\label{empiricalgrowthrate}
 \gamma_j\equiv \Delta \ln \textrm E_j[y_k]=\frac{\ln y_{jt^\prime}-\ln y_{jt}}{\Delta t}
\end{equation}
Assume $\Delta t=1$ for brevity. In the analysis in the main text, $\Delta t=5yrs$ as the data is analyzed in the 2014-19 span, but can be in principle any positive value.

To study growth in block groups where no aggregation has been performed, the empirical growth rate serves as the only growth quantity.
However, for studying the growth statistics of larger spatial units such as tracts and beyond, we have several choices. 
Consider the weights describing income share and population share
\begin{equation}\label{weightingschemes}
 \phi_{k|j}=\frac{p_{k}y_{k}}{\sum_{k^\prime\in j}p_{k^\prime}y_{k^\prime}}; \hspace{.25cm} f_{k|j}=\frac{p_k}{\sum_{k^\prime\in j} p_{k^\prime}}.
\end{equation}
The first weight defines an expectation value weighed additionally by income, whereas the second weight defines a standard population-weighted average.

Using these two weighting schemes, we now define the aggregation methods. The first option is to aggregate incomes according to Eq. \ref{aggincome}.
Then, the growth rate is computed
\begin{equation}\label{eg:oleypakessupp}
 \gamma_j^y\equiv \textrm E_j[ y_k\Delta\ln y_k]/\textrm E_j[y_k]=\sum_{k\in j}\phi_{k|j}\gamma_{k}
\end{equation} 
This measure explicitly considers the context of growth by factoring in the nominal income value.
This method is standard in the capital allocation literature \cite{olley1992dynamics}, and involves an expectation over the empirical growth rates, weighed by their income. 
Note that we explicitly weigh by the initial population composition, consistent with a {Laspeyres} aggregation scheme, whereas a {Paasche} aggregation would instead weigh by the {final population composition} \cite{Balk2008}.
The distinction is analogous to that between base-weighted and current-weighted index numbers in price and productivity measurement. Under the Laspeyres aggregation, aggregate growth reflects changes in incomes while holding the population structure fixed at its initial distribution, thereby isolating {within-unit} changes. In contrast, the Paasche formulation evaluates growth using the end-of-period composition, incorporating \textit{realized} reallocation or \textit{sorting} effects. It measures how shifts in population shares toward higher or lower income areas contribute to overall change.

In this paper, we adopt the Laspeyres specification so that compositional effects arising from population reweighting can later be identified explicitly through a Price decomposition.
Thus defines the second aggregation method.
We to first aggregate log income according to Eq. \ref{logincome}, an approach not standard in the economics literature, then compute the growth as a difference using the standard weights
\begin{equation}\label{meanloggrowth}
 \gamma_j^p\textrm =\Delta \textrm E_j[\ln y_k]=\sum_{k}\big[f_{k|j,t^\prime}z_{k|j,t^\prime}
-f_{k|j,t}z_{k|j,t}\big].
\end{equation}
Note that this explicitly re-weights the growth of each spatial unit by its population between time steps. As we will see later, this allows for an explicit retrieval of population sorting along incomes using the Price equation. 

These two calculations differ particularly by the order in which the logarithm and average are applied. 
When growth rates are small, we can apply the first order approximation $y_k\Delta\ln y_k\approx \Delta y_k$.
Applying this substitution this, we find that
\begin{equation*}
 \gamma^y_k\approx\frac{\textrm E[\Delta y_k]}{\textrm E[y_k]}\approx \Delta\ln\textrm E[y_k]=\gamma_k,
\end{equation*}
This final expression states that the income-weighted aggregagte growth rate approximates to the first order \ref{empiricalgrowthrate}.
This provides some theoretical justification for the close fit between empirical and income-weighted growth rates in Fig. \ref{fig:Fig1}C of the main text, and distinguishes the weighting schemes in Eq. \ref{weightingschemes} differ to higher order by Jensen's inequality.

\begin{figure*}\label{fig:App2}
 \centering
 \includegraphics[width=\linewidth]{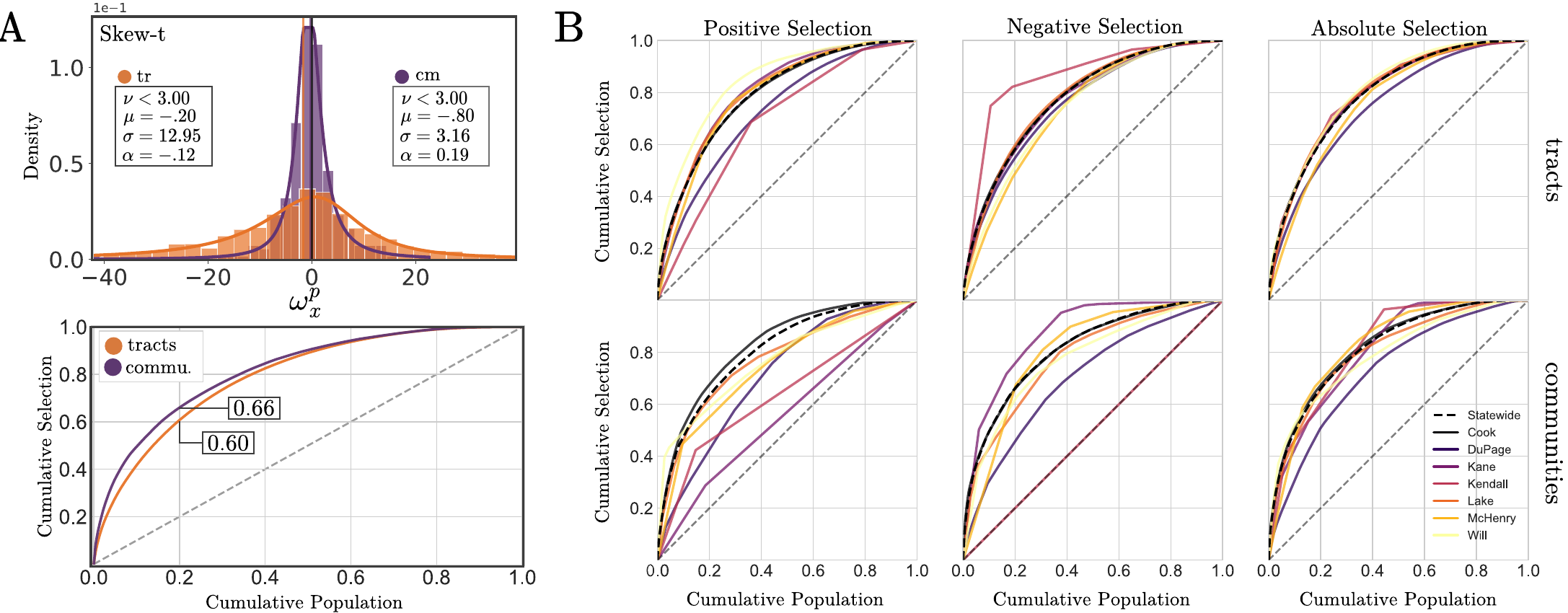}
 \caption{Income and population growth data is highly correlated and heterogeneous A) Selection effects within community areas (purple) and tracts (orange) are fat tailed and skewed, suggesting spatial correlations or spillover effects between units. Local heterogeneity and correlations in ACS income and population growth data at the tract level is lost when aggregating to community areas. Fat tails (above) match a high degree of concentration in a subset of the population (below). B) Aggregating income and log income reveal distributions reveal differing growth statistics. C) Differences between aggregated growth statistics and standard, empirical measures of growth encode information about growth dynamics.}

\end{figure*}

\section{Multiscale growth}

In this appendix, we define the price equation in terms of the growth rate and relative growth expressions derived in the previous sections. 

\subsection{Interpretation of selection}\label{selectioninterpretation}

Selection carries many meanings across disciplines. In urban sociology, it is typically measured at the individual level and reflects a combination of preferences, economic agency, and opportunity.
In population ecology, selection refers to differential success across heterogeneous traits under environmental pressures. Here, we interpret selection as drawing from both perspectives. Social and economic conditions within the city exert pressure on prospective and existing residents through employment and residential opportunities, competition, and prevailing costs. In this sense, the city is treated as an environment that both emerges from and selects for individuals’ economic capacity.
This selection encompases structural constraints such as housing costs, discrimination, zoning; analogous but distinct from segregation measures such as Theil's index and the multigroup index \textit{H}, which decompose a level of segregation into within- and between-unit components rather than the change in an aggregate quantity 

At the same time, the selection term measured in these decompositions should not be interpreted as a direct measure of individual choice or a causal estimate of sorting behavior. Rather, when applied to aggregated census data, the selection term can be interpreted as an aggregate signature of preference-linked sorting, but only insofar as preferences are expressed through the constraints and opportunities of the urban environment. Because U.S. census data do not provide direct access to individual preferences or longitudinal residential decisions, our interpretation is limited to an inference about the joint effect of underlying urban conditions and the aggregation of individual choices. Longitudinal data would provide more direct access to residential choice as an expression of individual-level selection preferences. Accordingly, the implementation in this work does not measure individual selection in the sense customary in urban sociological studies, nor does it by itself identify the specific mechanisms generating the observed compositional change.

\subsection{Price equation decomposition of log incomes}\label{section:pricedecomp}

For an average trait $z_j=\textrm E_ j[\ln y_k]$ , The price equation states that.

\begin{equation}\label{populationprice}
 \mathrm \Delta \textrm E_j[\ln y_k]=\tfrac{1}{w_j}\textrm{cov}[ w_k,z_k]+\tfrac{1}{w_j}\textrm E
 [w_kz_k
 ].
\end{equation}

\noindent If we assume $\Delta t=1$, it follows that $\Delta \textrm E[\ln y]_j=\gamma^p_j$. As explained in the main text, this expression decomposes the expectation value over growth rates (the average of log quantities) into compositional changes between units, the selection term, and endogenous shifts in that quantity in each unit (the transmission term). It is important to note that a log incone, $z_j$ is an average over log incomes, and explicitly not the log of an average over incomes. Most precisely, its written as

\begin{equation*}
z_j=\sum_{l\in k}f_{k|j}\textrm E_k[\ln y_l].
\end{equation*}
For the terminal child unit (in our case, the block group), $z_j=\ln y_j$
Assuming $k$ is not the terminal child, we can define the Price equation for each growth term in the transmission term as

\begin{equation}\label{populationpricelevel2}
 \gamma_k^p=\tfrac{1}{w_k}\textrm{cov}[w_l,z_l]+\tfrac{1}{w_\kappa}\textrm E
 [w_l\gamma_l^p].
\end{equation}
We can plug this expression into Eq \ref{populationprice} to retrieve

\begin{equation}
\gamma_j^p=\tfrac{1}{w_j}\textrm{cov}_j\big[w_k, z_k\big]+\tfrac{1}{w_j}\textrm E_j
 \big[\textrm{cov}_k[w_l,z_l]+\textrm E_k
 [w_l\gamma_l^p
 ]\big].
\end{equation}

More generally, the Price equation for the growth rate of any subunit
is $\gamma_{k}^p=T_{l\rightarrow k}^p+S_{l\rightarrow k}^p$.
We substitute this into the transmission term of Eq. \ref{pricemaintext} to retrieve the two-level Price decomposition as $\gamma_{j}^p=S_{k\rightarrow j}^p+S_{l\rightarrow j}^p+\textrm T_{l\rightarrow j}^p$, where an index spanning more than one level implies averaging $X_{l\rightarrow j}=\textrm E_j[X_{l\rightarrow k}]$. 
This redefines a previously endogenous transmission term $T_{k\rightarrow j}^p$ in Eq. \ref{pricemaintext} in terms of selection driven by internal sorting, $S_{l\rightarrow j}^p$, and transmission $T_{l\rightarrow j}^p$.
Thus, a multi-level decomposition is the weighted sum over the selection effects measured independently and locally at each level, transmitted to the aggregate level through population-weighted averaging of subunits. 

Now that we have a concise definition for the two-level decomposition, we can describe the decomposition for the dataset across multiple levels. Consider block groups, denoted $\textrm{bg}$, nested in tracts, denoted $\textrm{tr}$, nested in community areas, denoted $\textrm{cm}$, nested in counties, denoted $\textrm{ct}$, nested in the Illinois-side MSA, denoted $\textrm{m}$, we can define the decomposition for log incomes (via the Price equation) as

\begin{equation}
\begin{split}\label{priceequationmultilevel}
 \gamma_{\textrm{m}}^p=&\textrm E_{\textrm{m,ct,cm,tr}}\big[\bar w_{\textrm{bg}}\gamma_{\textrm{bg}}^p
 \big]+\textrm E_{\textrm{m,ct,cm}}
\Big[\textrm{cov}_{\textrm{tr}}\big(\bar w_{\textrm{bg}},z_{\textrm{bg}}\big)\Big]\\
&+
 \textrm E_{\textrm{m,ct}}
 \Big[\textrm{cov}_{\textrm{cm}}\big(\bar w_{\textrm{tr}},z_{\textrm{tr}}\big)\Big]+\textrm E_{\textrm{m}}
 \Big[\textrm{cov}_{\textrm{ct}}\big(\bar w_{\textrm{cm}},z_\textrm{cm}\big)\Big]\\
 &+\textrm{cov}_{\textrm{m}}\big(\bar w_{\textrm{ct}},z_{\textrm{ct}}\big).
\end{split}
\end{equation}
Here, growth at each scale is normalized to the MSA, such that $\bar w_j=w_j/w_{\textrm{m}}$.
This can be compactly written, as is in the main text,
 \begin{equation}
\begin{split}\label{priceequationmultilevelmaintext}
 \gamma_{\textrm{m}}^p=&S_{\textrm{ct}\rightarrow \textrm{m}}^p+S_{\textrm{cm}\rightarrow \textrm{m}}^p
 +
 S_{\textrm{tr}\rightarrow \textrm{m}}^p +S_{\textrm{bg}\rightarrow \textrm{m}}^p
 +T_{\textrm{bg}\rightarrow \textrm{m}}^p
 .
\end{split}
\end{equation}

A similar multiscale decomposition can be defined for Eq. \ref{eg:oleypakessupp}, where the decomposition is given by the Olley-Pakes decomposition
\begin{equation}
\gamma_j^y=\textrm E_j[\gamma_k^y]+\textrm{cov}_j[f_{y_k},\gamma_k^y].
\end{equation}
This quantity measures the relationship between incomes and growth rates, and is closely predicts first order contributions to income growth variance, a quantity which standard inequality measures like Gini is monotonic \cite{kemp2022statistical}.
The full decomposition is then
\begin{equation}
\begin{split}\label{priceequationmultilevel}
 \gamma_{\textrm{m}}^y=&\textrm E_{\textrm{m,ct,cm,tr}}\big[\gamma_{\textrm{bg}}^y
 \big]+\textrm E_{\textrm{m,ct,cm}}
\Big[\textrm{cov}_\textrm{tr}[f_{y_\textrm{bg}},\gamma_\textrm{bg}^y]\Big]\\
&+
 \textrm E_{\textrm{m,ct}}
 \Big[\textrm{cov}_\textrm{cm}[f_{y_\textrm{tr}},\gamma_\textrm{tr}^y]\Big]+\textrm E_{\textrm{m}}
 \Big[\textrm{cov}_\textrm{ct}[f_{y_\textrm{cm}},\gamma_\textrm{cm}^y]\Big]\\
 &+\textrm{cov}_{\textrm{m}}[f_{y_\textrm{ct}},\gamma_\textrm{ct}^y].
\end{split}
\end{equation}
We analyze the relationship between income selection effects and changes in the Gini index for spatial, computed is $\Delta G_{\textrm{tr}\rightarrow \textrm{cm}}=G_{t^\prime}-G_t$ 
The Gini index is computed for a spatial unit by applying the standard Lorenz curve method to the child units' income data at the initial and final timestep.
We find that selection predicts change in Gini, that is more positive selection predicts increases to inequality, with an $r^2=.5$ across communities and $r^2=.42$ across tracts.

We also partitioned the Gini change data to compare predictions of increases and decreases in Gini. Across communities, selection predicts positive Gini change, $\Delta_\textrm{cm}^{z+}$ better with $r^2=.6$ than negative Gini change, $\Delta_\textrm{cm}^{z-}$ with $r^2=.45$.
\begin{figure}[h]
 \centering
 \includegraphics[width=\linewidth]{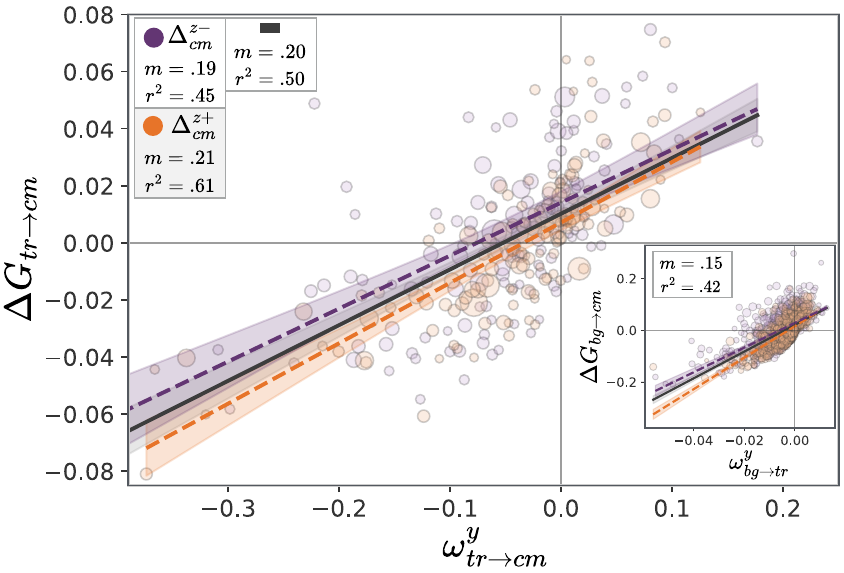}
 \caption{Income selection effects, $\omega_j^y$ predict changes in gini within communities and tracts (inset). The spatial dynamics of income-selection will be explored in future work.}\label{fig:App2_5selection}
\end{figure}

\begin{figure}[h]
 \centering
 \includegraphics[width=.75\linewidth]{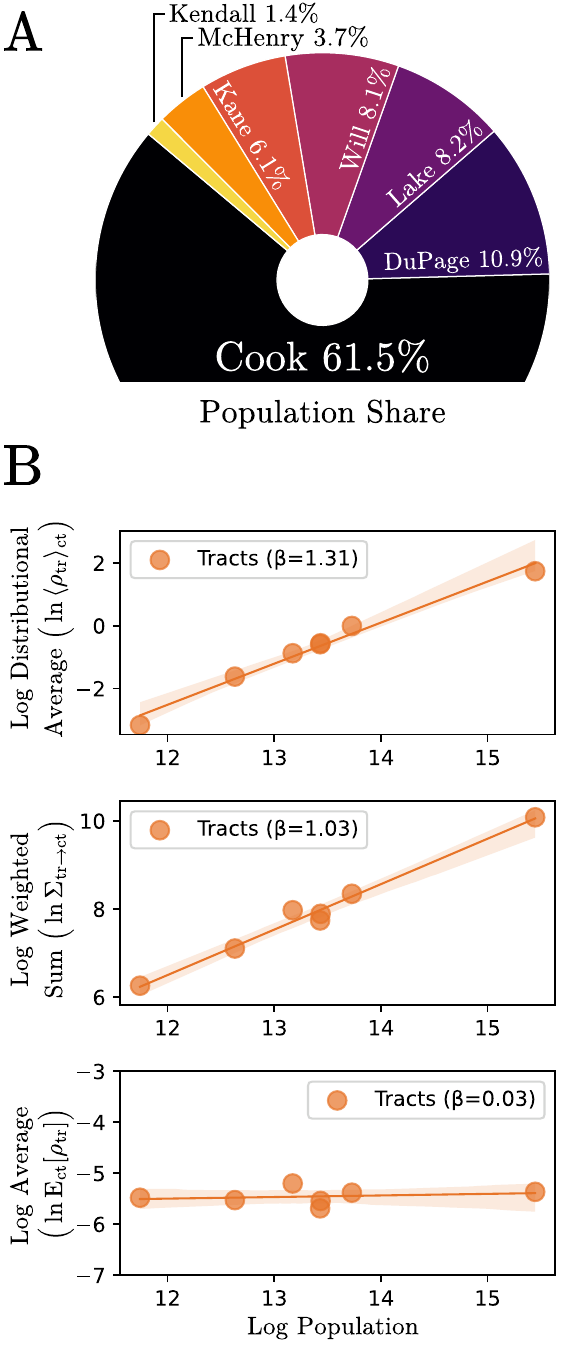}
 \caption{Scaling effects are observed in the data. A. relative population shares by county. B. Selection scales superlinearly across counties (top), however this becomes linear when accounting for relative population between tracts (middle), and scale invariant when accounting for tract and county population differences. }\label{fig:App3}
\end{figure}

\section{Skew-t Distribution}

In this section we demonstrate a skew-t distribution, which fits selection data well. A Student-t distribution behaves like a Gaussian distribution with fatter tails. 
The Student-t distribution is defined

\begin{equation}
 p^\prime(x|v)=\frac{\Gamma(\frac{\nu+1}{2})}{\sqrt{v\pi}\Gamma(\frac{\nu}{2})}\bigg(1+\frac{x}{\nu^2}\bigg)^{\frac{\nu-1}{2}},
\end{equation}
where the domain is defined over all real $x$ with degrees of freedom (DOF) $\nu$. 
The gamma function is given by $\Gamma(\cdot)$, a function often used to normalize gamma distributions. 
The DOF is bounded below by zero, and the tails converge to that of a Gaussian (shrink) as $\nu\rightarrow\infty$.
In this limit, the ratio of gamma functions converge to $(\tfrac{\nu}{2})^{1/2}$, and the prefactor converges to $\tfrac{1}{\sqrt{2\pi}}$

The mean of a Student-t distribution can be shifted with parameter $\mu$, and the variance scaled with $\sigma$. 
The shifted Student-t becomes
\begin{equation}
 p(x|\nu,\mu,\sigma)=\frac{1}{\sigma}p^\prime\bigg(\frac{x-\mu}{\sigma}|\nu\bigg),
\end{equation}
which converges to a Gaussian centered about $\mu$ and a standard deviation of $\sigma$ as $\nu\rightarrow\infty$.

The Skew-t distribution is then modified by the skew parameter $\alpha$, which is right skewed (the right tail is fatter) when $\alpha>0$ and left skewed with $\alpha<0$.
The full definition of the Skew-t is
\begin{equation}\label{eq:fulldistribution}
 p(x|\nu,\mu,\sigma,\alpha)=\frac{2}{\sigma}p^\prime(z|\mu))P^\prime\bigg(\alpha z\sqrt{\frac{\nu+1}{\nu+z^2}}|\nu+1\bigg),
\end{equation}
where the Gaussian standardized value is defined $z=\tfrac{x-\mu}{\sigma}$. The skew is given by the CDF of $p^\prime$, and is defined 
\begin{equation}
 P^\prime(x|a)=\int_{-\infty}^xp^\prime(u;a)du,
\end{equation}
which evaluates to a cumbersome expression that will not be fully defined here.
This is a modulating function that adds weight to the Student-t distribution depending on the integration limit.
When the skew factor $\alpha=0$, the function evaluates to 0.5, and the factor of 2 in equation \ref{eq:fulldistribution} cancels, reducing to the symmetric Student-t distribution. When $\alpha \neq 0$, the CDF creates asymmetry by differentially weighting the left and right tails of the distribution.
For $\alpha > 0$, positive values of $z$ produce larger integration limits, causing $P^\prime$ to exceed 0.5 and amplify the right tail. Conversely, negative values of $z$ produce smaller integration limits, causing $P^\prime$ to fall below 0.5 and suppress the left tail. This creates a right-skewed distribution with a longer tail extending to positive values. 
This construction allows the skew-t distribution to model data exhibiting both heavy tails (controlled by $\nu$) and asymmetry (controlled by $\alpha$), making it particularly suitable for real-world phenomena such as distributions of selection values where correlated local fluctuations cause extreme values to occur more frequently than predicted by normal distributions.

The fit for the tract level selection data has higher skew, $\alpha_{\textrm{tr}}=0.55$, than community level, $\alpha_{\textrm{cm}}=0.20$, with fatter tails as measured by the degrees of freedom parameter, $\nu_{\textrm{tr}}=1.43$ to $\nu_{\textrm{cm}}=2.47$.

\section{Scaling Behavior of Selection}\label{scaling}

Urban scaling theory posits that many properties of cities can be described by a simple power-law relationship with their population, $p$, $S=S_0p^\beta$. Here, $S$ is the observed, urban indicator while $\beta$ is the scaling exponent.
When expressed on a logarithmic scale, the expression takes a linear form 
\begin{equation}
\ln S=\ln S_0+\beta\ln p.
\end{equation}
We see clearly that $\beta$ describes how the indicator changes with city size.
Analyzing its value gives us insights into potential mechanisms for how people interact, move and self sort around cities \cite{bettencourt2013origins}.

In this section, we will discuss the scaling behavior of the empirical selection quantities, aggregated to the county level.
By comparing relative selection quantities between counties, we gain some insights on the spatial assortment and scale-dependence of sorting.
We note that this is not a typical application of scaling theory, for which the city is the unit of measurement and not subdivisons of the city and analysis is conducted cross-sectionally across an ensemble of cities. 
As such, the following results cannot be interpreted under standard scaling theory. 
Secondly, as this analysis is performed for a handful of counties in a single city, we do not attempt to extrapolate these results nor make any general claim about how selection scales in cities or subunits of cities. 
Instead, we only aim to interpret the scaling results Chicago dataset, and identify patterns worth exploring across other datasets.

In Fig. \ref{fig:App3}, we establish the population share of Chicago MSA counties in Illinois. 
Cook county, in which the City of Chicago and other urban centers, like Evanston and Oak Park are location, accounts for nearly 2/3rds of the MSA population. 
Other population centers like Aurora and Naperville, are internally split between the counties of DuPage, Lake, and Will and so on. 
These populationn splits will become important as we re-weigh selection effects in the scaling analysis.

The first quantity we report is a distributional average selection selection scaling effect, reported at the top of in Fig. \ref{fig:App3}B.
Here, 
\begin{equation}
\langle\rho_\textrm{tr}\rangle_\textrm{ct}=\frac{\sum_{\textrm{tr}\in\textrm{ct}}|S_\textrm{tr}^p|}{N_{\textrm{tr}}^{\textrm{ct}}},
\end{equation}where $N_\textrm{tr}^\textrm{ct}$ counts the number of tracts in the county. 
This quantity measures the total geographic churn, treating a selection event in a low-density tract the same as one in a high-density tract. It measures the total magnitude of selection events across all places.
In the data, we observe a scaling coefficient of $\beta_d=1.31$,
indicating county that is 10\% larger in population has approximately 13\% more total selection magnitude across its tracts
The superlinearity indicates that selections effects are disproportionately more intense in larger counties.

This measurement equally weighs each spatial unit. 
If we account for spatial population heterogeneity across units within county, we can measure more directly the human impact on selection. 
We will do this in two steps: first by just considering the apportionment of households internal to each county.
The second will introduce weighing counties relative population, providing a per-capita measurement of selection. 

The population-weighted selection is computed as
\begin{equation}
\Sigma_{\textrm{tr}\rightarrow\textrm{ct}}=\sum_{\textrm{tr}\in\textrm{ct}}p_\textrm{tr}|S_{\textrm{tr}}^p|,
\end{equation}
and measurs the total magnitude of selection as experienced by the population.
We measure a virtually linear scaling coefficient of $\beta_w=1.03$. 
This means that selection across units scale superlinear with county size, the total impact on the population scales directly with size.
This implies that the more intense sorting events observed in larger counties are preferentially occurring in less populated tracts. This frontier of demographic change in Chicago lies in more sparsely populated populated tracts across counties.

Finally, when we consider the relative population of each county, as we do for the expression
\begin{equation}
\textrm E_\textrm{ct}[\rho_\textrm{tr}]=\frac{\Sigma_{\textrm{tr}\rightarrow\textrm{ct}}}{p_{ct}}.
\end{equation}
This expression measures the average, per-capita intensity of selection.
We observe a $\beta_p=.03$, meaning that for Chicago, experienced selection is scale-invariant.
This confirms the insight from before, that an average person in a large county experiences the same intensity of selection as an average person in a small county. 

Generally, we observe that while the system as a whole produces more total churn as indicated by the superlinearity of $\beta_d$, it has organized itself in such a way that the average experience remains constant. It implies that the sorting processes are not accelerating on a per-person basis with city size, and that in this limited case, the "social accelerator" effect that makes cities more productive or innovative does not seem to apply to the intensity of neighborhood churn.

\end{document}